# Comparing Different Uncertainty Measures to Quantify Measurement Uncertainties in High School Science Experiments

**Karel Kok \***
Humboldt-Universität zu Berlin
karel.kok@physik.hu-berlin.de

**Burkhard Priemer**
Humboldt-Universität zu Berlin
priemer@physik.hu-berlin.de

**Abstract**
Interpreting experimental data in high school experiments can be a difficult task for students, especially when there is large variation in the data. At the same time, calculating the standard deviation poses a challenge for students. In this article, we look at alternative uncertainty measures to describe the variation in data sets. A comparison is done in terms of mathematical complexity and statistical quality. The determination of mathematical complexity is based on different mathematics curricula. The statistical quality is determined using a Monte Carlo simulation in which these uncertainty measures are compared to the standard deviation. Results indicate that an increase in complexity goes hand in hand with quality. Additionally, we propose a sequence of these uncertainty measures with increasing mathematical complexity and increasing quality. As such, this work provides a theoretical background to implement uncertainty measures suitable for different educational levels.

**State of the literature**
- Judging the quality of data and interpreting data is a competence that is growing in importance in everyday life.
- Students experience a lot of difficulties with the interpretation of measurement uncertainties.
- Calculating the standard deviation proves difficult but an automated process of calculating it takes the focus away from the interpretation of the result.

**Contribution of this paper to the literature**
- This article presents several alternative quantitative measures to express the measurement uncertainty (measure of spread).
- These measures are examined in terms of mathematical complexity and statistical quality.
- A sequencing of these measures is presented to help introduce the topic of measurement uncertainties, leaving the focus on its interpretation.

## 1. Introduction

A measurement uncertainty quantifies the doubt about the validity of an experimental result and is an indication of the quality of the data. Additionally, it allows for a comparison of experimental results or models within a certain degree of confidence. In today's society, being able to judge the quality of data and making decisions based on it, is a skill that is gaining more and more importance (Chinn & Malhotra, 2002; Holmes et al., 2015; Sharma, 2006). However, the topic of measurement uncertainties is rarely addressed at high school level (Priemer & Hellwig, 2018; Möhrke, 2020). Often, the calculation of the uncertainty (most often the standard deviation) proves difficult for students and consequently takes the focus away from the interpretation of the result (Zangl & Hoermaier, 2017). Priemer and Hellwig (2018) suggest the total span of the measurements as an uncertainty interval. This simplifies the calculation and allows for more focus on the interpretation. The downside is that this measure is highly susceptible to outliers.

We conceptualize the standard deviation and the total span of measurements as two extremes in a range of possibilities to describe the variation of measurement data. In this article, we look at the usability of several alternative uncertainty measures which are based on existing measures of spread. These measures will be easier to determine than the standard deviation, in terms of mathematical complexity, but yield better quality results than the total span. We base the complexity on the mathematics curricula and the quality on the results of our Monte Carlo simulation. Our work aims to build a foundation to introduce quantitative measures for measurement uncertainties in education. We analyze their properties and discuss advantages, disadvantages, and limitations from a didactical perspective. Thus, we argue, that our results are a necessary prerequisite to develop educational reconstructions, teaching material, or empirical studies regarding measurement uncertainties.

We start with an overview of the literature regarding measurement uncertainties within science and statistics education research in section 2. The alternative uncertainty measures are presented in section 3. Section 4 describes the Monte Carlo simulation in which a main set of measurements is generated. The alternative uncertainty measures are then calculated from subsamples and compared with the standard deviation of the main set. The criteria for usability are described in Section 5. Results are presented in section 6 and discussed in section 7.

## 2. Theoretical Framework

Working with variation in data can be seen as a core practice of statistical thinking. Moore (1990) describes statistical thinking as an awareness of the omnipresence of variation in processes, the need for data, an awareness of the sources of variation, the quantification of variation, and the explanation of variation. Given that students experience difficulties in understanding basic statistical quantities like variance (Garfield & Ben-Zvi, 2007; Torok & Watson, 2000) it becomes obvious that they face similar difficulties with the meaning of measurement uncertainties (Kok et al., 2019; Lubben et al., 2001; Séré et al., 1993). Because measurement uncertainties are mathematically expressed by measures of spread, knowledge about statistics is necessary to understand the meaning and implication of measurement uncertainties. However, it is not sufficient. Buffler et al., (2001, p.1151) indicated that the ability to calculate means and uncertainties does not imply an understanding of the statistical nature of these measurements. Even if technical obstacles in estimating uncertainties are eliminated it is still a challenge for students to grasp the meaning of uncertainties. Zangl and Hoermaier (2017) developed a digital tool for students to calculate the mean, standard deviation, and propagation of uncertainties. This shifted university students' attention from calculating the uncertainty to the interpretation of the result. However, the disadvantage they noted was that, because of the automated calculation, students appear to lose the feeling for whether or not the outcome is of the correct magnitude. This shows that there is an interplay between the formal statistics calculations and the interpretation of measurement uncertainties. A notion supported by Casleton et al., (2014), who confirmed a strong link between the understanding of measurement uncertainties and the appreciation of the value of statistics.

### 2.1 Students' Difficulties in Understanding the Concept of Variation

The subject matter model by Priemer and Hellwig (2018), describes the contents needed to call attention to the relevance of measurement uncertainties in secondary education. Although the model mentions a measure of spread, the model is independent of this specific measure, mostly because the interpretation of the measurement result is independent of the chosen measure of spread (see also Torok & Watson, 2000, p.166). The model can be used to identify students' difficulties in understanding measurement uncertainties. Here, we focus on the dimension "assessment of uncertainties" which includes the concept of variation.

As described in the Guide to the Expression of Uncertainty in Measurement (GUM, Joint Committee for Guides in Metrology, 2008), the uncertainty in a type A analysis –i.e., the statistical analysis of repeated measurements– follows from the probabilistic distribution of these measurements. As the repetition of measurements is a means to see this distribution, it is no surprise that students on different educational levels find it hard to perceive the need for repeated measurements (Allie et al., 1998; Lubben & Millar, 1996; Séré et al., 1993). They often regard the repetition of measurements as a means of practicing one's measurement skill or as a verification of previous measurements (Buffler et al., 2001). This corresponds to what Lubben et al., (2001) have called the point-paradigm. In the point paradigm, students regard measurements as isolated events and the mean value as the "true" outcome of the experiment. Whereas in the set-paradigm, students regard the measurement series as a whole, indicating a range in which the measured quantity can be expected.

However, a set-based argumentation with data seems to pose problems for students. In a study 10–14 years old students were asked to investigate the (non-)covariation of variables in an experiment (Kanari and Millar, 2004). It was found that students experience difficulties, especially in the non-covarying cases, due to a lack of understanding of measurement uncertainty.

Lubben and Millar (1996) give a good summary of the different views students have on repeated measurements and variation. They indicate eight levels of students' understanding of the collection and evaluation of empirical data that range from "measure once and you will get the right value" to repeated measurements to get an indication of the best estimator and its spread.

### 2.2 Approaches in Teaching the Concept of Variation

Despite the difficulties that students have in understanding measurement uncertainties, there are indications that students can reason with and identify (components of) uncertainties as early as third grade (Gal et al., 1989; Masnick & Klahr, 2003; Metz, 2004; Petrosino et al., 2003).

Ford (2005), Munier et al., (2013), Pols et al., (2019) report lively discussions where students are encouraged to think about the credibility and limitations of their results. In all cases, students realize that reporting an indication of the uncertainty will be necessary. In the case of Munier et al., (2013), this is even done in a quantitative manner.

The above examples are encouraging as they show that young students are, with help, able to see the necessity of discussing measurement uncertainties as well as identify (sources of) measurement uncertainties. What these studies lack, however, is an agreed-upon quantification of the uncertainty. This would allow the students to compare their results and judge the quality of their experiment.

Garfield and Ben-Zvi (2005) have constructed a framework for teaching and assessing reasoning about variability. Although this framework mentions different measures of variability (standard deviation, interquartile range, and range), it is limited in the sense that the differences are stated by mentioning what they should measure. No assessment is made about the quality of these measures in situations with small sample sizes —common for high school experiments.

In the science education literature, we have found no research-based recommendations for using measures of spread suitable to express measurement uncertainties, despite the necessity to teach about measurement uncertainties in schools. From this, we conclude that there is a need for practical uncertainty measures that have a fitting complexity and yield good quality results. With complexity, we mean the mathematical complexity and the ease of calculation. With quality, we mean how well the uncertainty measure describes the variation of the measurements as compared to the standard deviation. The next section lists the uncertainty measures –all based on existing measures of spread– that we have selected for our comparison.

In this article, we interpret all measures of spread as quantifications of measurement uncertainties. Therefore, we use the terms uncertainty measure and measure of spread synonymously.

## 3. Uncertainty Measures

The calculation of each of the uncertainty measures (i.e., quantified measure of spread) starts with the calculation of the arithmetic mean:

$$\bar{x} = \frac{1}{N}\sum_i x_i \tag{1}$$

where $x_i$ is one measurement, and $N$ is the number of repeated measurements.

### Min-Max

The min-max uncertainty is an adaptation of the range, which is the total span of the measurement series (Barlow, 1993, p.12). To make this range symmetric, we take the uncertainty to be the largest distance between the minimum or maximum of the measurement series as compared to the mean, similar to what is suggested by Priemer and Hellwig (2018).

We do this by first sorting the samples from small to large: $\{x_1, ..., x_N\}$ and taking the maximum value of the mean minus the smallest value and the largest value minus the mean:

$$u_{\text{min-max}} = \max(\bar{x} - x_1, x_N - \bar{x}). \tag{2}$$

### Exclude Extremes

The procedure for this uncertainty measure is the same as the min-max uncertainty, except that the minimum and maximum value of the measurement series are excluded in the determination of the uncertainty, reducing the influence of outliers. The uncertainty is then calculated using the set $\{x_2, ..., x_{N-1}\}$:

$$u_{\text{excl.extr.}} = \max(\bar{x} - x_2, x_{N-1} - \bar{x}). \tag{3}$$

### Middle 50%

To calculate the uncertainty for the middle 50%, we again sort the measurements from small to large. Then, the minimum and maximum values are excluded one pair at a time, as long as the remaining measurements constitute to 50 % or more of the original measurements. The uncertainty is then calculated using the set $\{x_{N/4}, ..., x_{N-N/4}\}$ where $N/4$ will always be rounded down:

$$u_{\text{middle50\%}} = \max(\bar{x} - x_{(N/4)+1}, x_{N-N/4} - \bar{x}). \tag{4}$$

### Mean Absolute Deviation

The mean absolute deviation (MAD, see e.g., Barlow, 1993) is calculated by taking the mean of the absolute values of the difference between the measurements and the mean value:

$$u_{\text{MAD}} = \frac{1}{N}\sum_i |x_i - \bar{x}|. \tag{5}$$

### Standard Deviation

Lastly, we can calculate the sample standard deviation of the measurement series:

$$u_{\text{SD}} = \sigma_x = \sqrt{\frac{\sum_i (x_i - \bar{x})^2}{N-1}}. \tag{6}$$

### 3.1 Other Uncertainty Measures

We have limited ourselves to the previous uncertainty measures. There are of course other (established) measures of spread that could have been used to create an uncertainty measure. For instance, the interquartile range (IQR, see e.g., Barlow, 1993) or discarding different percentages of data.

In practice, these and other measures yielded no better results than our selection of alternative uncertainty measures, as we will show in the discussion in section 7.1. For the sake of clarity, we will constrain ourselves to the above measures in our analysis.

## 4. Monte Carlo Simulation

The Monte Carlo method is a computational method where the numerical simulation of an experiment is repeatedly executed, using input parameters with random values from a given probability distribution. It can be used to simulate experiments thousands of times to probe the effect of the different input parameters but also to probe measurement uncertainties in experiments (Brandt, 2014). We will use the Monte Carlo method, programmed in Python 2.7.15, to calculate the alternative uncertainty measures and probe their behavior (see supplemental material).

We start the simulation by generating a random set of $N_{\text{set}} = 1,000$ normally distributed values. This set represents measurement data that could have been obtained in an experiment if the measurements were repeated 1,000 times. Assuming the distribution to be a normal distribution is valid under the central limit theorem (in short: the sum of $x_i$ becomes a normal distribution for large $N$, see also Cowan, 1998, p. 147). We assume this to hold for experiments in an educational setting, even with a small number of measurements (Taylor, 1997). As such, our set consists of 1,000 normally distributed values with a known mean $\mu_{\text{set}}$ and standard deviation $\sigma_{\text{set}}$. This standard deviation will be the reference value with which we shall compare the alternative uncertainty measures.

Next, we take a random subsample size with an initial value of $N = 4$ values from this set –representing four repeated measurements of an experiment. Using this subsample, we calculate the values of each of the alternative uncertainty measures.

To compare the alternative uncertainty measures with the standard deviation of the set (being the reference value describing the spread of the data), we define the *uncertainty deviation*: the difference between the alternative uncertainty measure of the subsample and the standard deviation of the complete set $\sigma_{set}$ divided by the standard deviation of the complete set:

$$\Delta_i \equiv \frac{u_i - \sigma_{set}}{\sigma_{set}} \quad (7)$$

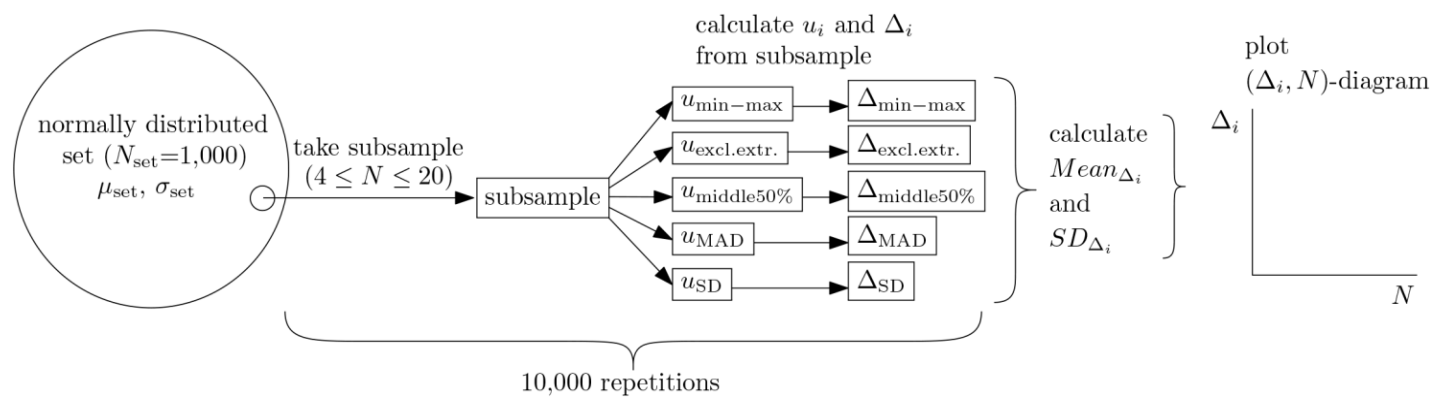

**Figure 1.** A schematic flow-chart of the Monte Carlo simulation.

where $u_i$ is one of the alternative uncertainty measures calculated from the subsample. Note that by doing this, the uncertainty deviation is independent of the values of the set parameters: $\mu_{set}$ and $\sigma_{set}$. Consequently, the results of this simulation are valid for all normal distributions.

The process of taking subsamples and calculating uncertainty deviations is repeated 10,000 times. What follows is a distribution of uncertainty deviations for the alternative uncertainty measures, from which we calculate the means and standard deviations.

We repeat the procedure for subsample sizes in the range of $N = [4; 20]$, typical numbers of repeated measurements in school experiments.

We then plot the mean value of the uncertainty deviation and its standard deviation as a function of $N$. This shows how the different uncertainty measures develop for different values of $N$ and indicate the quality of these uncertainty measures.

The whole procedure of the Monte Carlo simulation is summarized in **Figure 1**.

**Figure 2** shows the development of the mean value of the uncertainty deviation as a function of the number of repetitions for our alternative uncertainty measures, using a subsample size of $N = 10$, a common number of repetitions for high school experiments. From the figure, it can be seen that this mean uncertainty deviation stabilizes after 1,000 repetitions, indicating that our 10,000 repetitions will more than suffice.

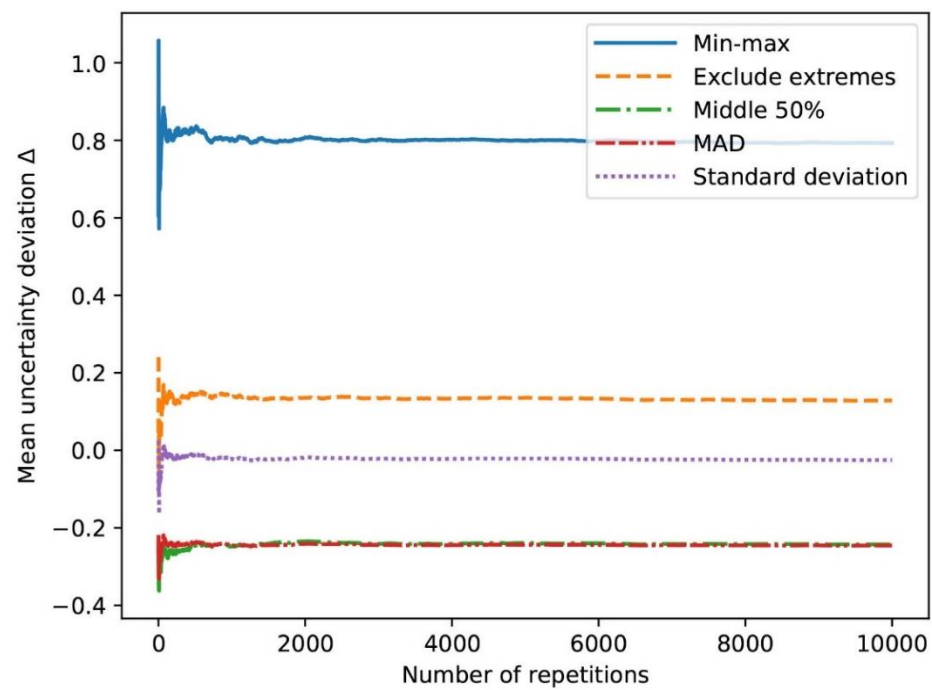

**Figure 2.** The development of the mean uncertainty deviations as a function of the number of repetitions, for a subsample size of $N = 10$. The values stabilize after 1,000 repetitions.

## 5. Classification
We classify the usability of the alternative uncertainty measures based on two categories: their *complexity*, i.e., the mathematical complexity in the calculation and their *quality*, how well the result describes the variance in the data.

### 5.1 Complexity
As described by Zangl and Hoermaier (2017), the more emphasis students put on the calculation, the less emphasis they put on the interpretation of the result. We, therefore, suggest arranging the alternative uncertainty measures based on the complexity of the mathematical operations.

The Common Core Standards for Mathematics (USA), Rahmenlehrplan Sek I Mathematik (GER), and the Kerndoelen Wiskunde (NLD) all show a similar structure for the middle school mathematics curricula (National Governors Association Center for Best Practices, 2010; LISUM, 2018; van der Zwaart, 2007). They all start with addition and subtraction around first grade, continue with multiplication of small numbers after third grade, and introduce the square root around eighth grade. The mean, median, modus, and absolute values are introduced around sixth grade. The Common Core Standards for Mathematics also mentions the IQR and MAD as a means to describe variability in data in sixth grade (p. 39). The standard deviation, however, is introduced at the high school level (p. 81).

Based on the sequencing in these mathematics curricula, we classify the complexity of the uncertainty measures based on mathematical operations: sorting, addition and subtraction, multiplication and division, absolute value, and squaring and rooting. The more advanced the mathematical operations, the higher the complexity of the uncertainty measure.

### 5.2 Quality
The quality of the uncertainty measure depends on how well this measure describes the variability of the data. The quality of an estimator is usually expressed in terms of bias, consistency, and efficiency and usually in that order (see e.g., James et al., 2007, p.128; Cowan, 1998, p.65; Barlow, 1993, p.68).

Bias indicates the difference between the expected value of an estimator $\langle \hat{a} \rangle$ and the value of what is being estimated $a$. The bias is calculated as: $b = \langle \hat{a} \rangle - a$. Consistency indicates whether, with increasing values of $N$, the estimator $\hat{a}$ approaches the estimated value: $\lim_{N \to \infty} \langle \hat{a} \rangle = a$. The efficiency is a measure of the variance of an estimator.

In this case, the estimators $\hat{a}$ are the alternative uncertainty measures $u_i$. The value we are trying to estimate $a$ is the standard deviation of the set $\sigma_{set}$. As such, the bias is given by the uncertainty deviation $\Delta_i$ but in a general form, as it is expressed in standard deviations (see equation (7)). We call this parameter the *deviation bias*. Because the uncertainty deviations will change for different values of $N$, we compare the deviation bias in the limit of $N \to \infty$. When the value of the deviation bias fails to stabilize to a consistent value, the uncertainty measure will be labeled: biased.

To describe the consistency, we use two derived quality parameters: *convergence* and *smoothness*. Convergence describes whether our alternative uncertainty measures will, for larger values of $N$, converge to a certain value. Due to deviation bias, this value does not necessarily need to be the standard deviation of the set $\sigma_{set}$.

The smoothness parameter indicates whether the uncertainty $u_i$ in the $(\Delta, N)$–diagram is smooth or that its development shows sharp positive and negative "kinks". The latter would indicate that for an increase in $N$, the behavior of $u_i$ could radically change, which is an undesired property.

Lastly, we describe the efficiency in terms of the *spread*. We quantify this spread as the standard deviation of the uncertainty deviation $SD_{\Delta_i}$ for a subsample size of $N = 10$. This value of $N$ is chosen because it is a common number of repeated measurements in high school experiments.

## 6. Results
**Figure 3** shows the mean uncertainty deviations for all alternative uncertainty measures as a function of $N$. The highlighted regions indicate the standard deviation of the individual values of $u_i$ (i.e., not the standard deviation of the mean). The black line at $\Delta = 0$ indicates the standard deviation of the set.

From this figure, we can see that the standard deviation, unsurprisingly, is the best approximation of the standard deviation of the set for all values of $N$. The mean uncertainty deviation quickly converges towards the black line and the uncertainty of the uncertainty deviation also decreases with increasing values of $N$.

The MAD is systematically lower than the standard deviation. This originates from the definition of the MAD, which takes the absolute deviations of the mean instead of taking the root of the sum of squares. The mean uncertainty deviation also quickly converges to a mean deviation of approximately $-.23$, and the uncertainty decreases with increasing values of $N$.

The middle 50% shows sharp kinks whenever $N$ is a multiple of four, corresponding to another pair of minimum-maximum-values to be excluded. For higher values of $N$, middle 50% converges towards the MAD.

As expected, the min-max uncertainty blows up quickly for larger values of $N$. Because we take the uncertainty to be the largest distance to the mean using the complete subsample, this measure is highly susceptible to outliers, which also explains its high standard deviation.

Exclude extremes starts off in the same fashion as the middle 50%, as they use the same measurements and uncertainty calculation in the range of $N = [4; 7]$. At high values of $N$, exclude extremes starts converging towards the min-max. Between subsample sizes $N = 8$ and $N = 9$, exclude extremes crosses the zero-uncertainty deviation line, indicating –on average– almost perfect agreement with the standard deviation of the set.

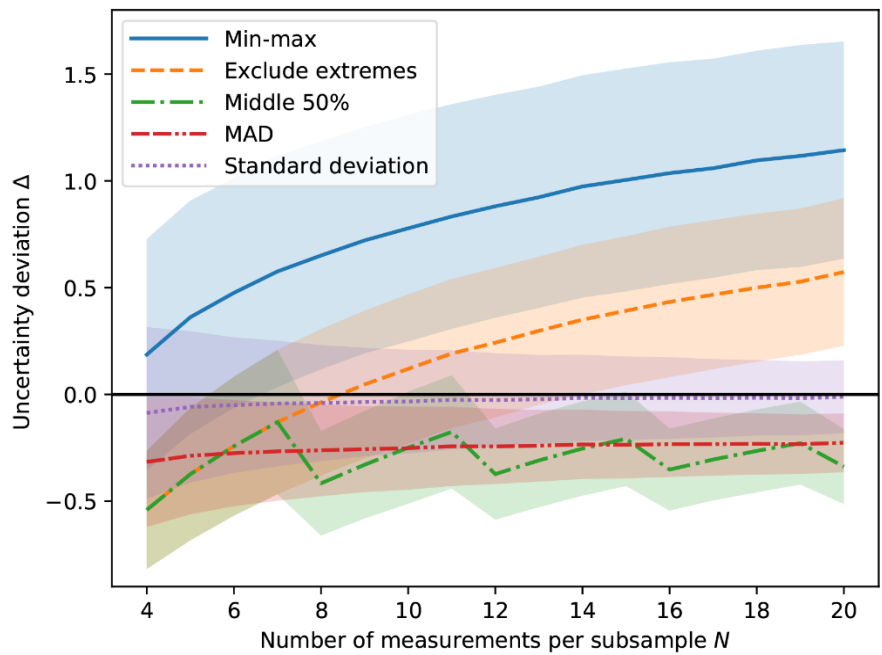

**Figure 3.** The mean values (lines) and standard deviations (shaded areas) of the uncertainty deviations of the different alternative uncertainty measures $u_i$ for different subsample sizes $N$. The horizontal black line indicates zero uncertainty deviation, corresponding to the standard deviation of the set.

The classification of the results in terms of complexity and quality as defined in section 5 is shown in **Table 1**. The uncertainty measures are sorted top to bottom from lowest to highest complexity.

Simultaneously the quality in terms of deviation bias, convergence, and spread (with the exception of the standard deviation) appears to increase along with the complexity. The min-max and exclude extremes measures are biased, the middle 50% and MAD each have a limiting value for the deviation bias, and the standard deviation has zero bias.

The only uncertainty measure with a non-smooth behavior is, by design, the middle 50% measure.

The spread of all measures decreases with increasing complexity (see the last column of **Table 1**). With a remarkable behavior for the MAD, which has a spread even lower than the standard deviation. This is due to the linearity of the terms in the calculation as compared to the standard deviation which uses squared terms.

**Table 1.** The classification of the uncertainty measures in terms of complexity and quality.

| $u_i$ | Complexity | | | | | Quality | | | |
|---|---|---|---|---|---|---|---|---|---|
| | sort | add, subtract | multiply, divide [a] | abs. | square, root | dev. bias | conv. | smooth | spread |
| Min-max | YES | YES | NO | NO | NO | biased | NO | YES | .52 |
| Exclude extremes | YES | YES | NO | NO | NO | biased [b] | NO | YES | .35 |
| Middle 50% | YES | YES | NO | NO | NO | −.34 | YES | NO | .26 |
| MAD | NO | YES | YES | YES | NO | −.21 | YES | YES | .19 |
| Standard deviation | NO | YES | YES | NO | YES | 0 | YES | YES | .24 |

[a] Although division is needed in the calculation of the arithmetic mean, this column only indicates the necessity for multiplication and/or division in the calculation of the uncertainty.
[b] Deviation bias = −.03 for a subsample size of $N = 8$, and .05 for a subsample size of $N = 9$.

## 7. Discussion
Our results suggest a sequencing of the uncertainty measures for science education, which we will discuss after briefly revisiting previously disregarded uncertainty measures.

### 7.1 Disregarded Measures
Looking at **Figure 3**, we see that the exclude extremes crosses the $\Delta = 0$ line between $N = 8$ and $N = 9$, yielding a similar result as the standard deviation. This behavior is, however, not very surprising. For exclude extremes, the uncertainty for subsample sizes $N = 8$ and $9$ is calculated using $75\%$ and $78\%$ of the measurements. This comes close to the $68\%$ of measurements that one standard deviation includes.

We further explore this characteristic by adapting the middle 50% and set the percentage to $68\%$ (similar to the standard deviation) and $76\%$ (the optimum percentage as extrapolated from the exclude extremes line in **Figure 3**).

To come as close as possible to this percentage, we will exclude values one by one until the number of remaining values in the subset divided by the original subsample size best approximates the chosen percentage. The values that are excluded are the values that are farthest from the mean (i.e., the largest value of $|x_i - \bar{x}|$). The uncertainty is calculated as the largest distance from the mean of the remaining subset, identical to the middle 50% procedure.

The last measure we compare, is the interquartile range (IQR), the range between the upper and lower quartiles. Where the quartiles are the medians of the upper and lower half of the sorted measurements. This measure of spread can be made into a symmetric uncertainty measure by taking the largest of the two distances from the quartiles to the mean.

**Figure 4** shows that the 76% of measurements oscillates around $\Delta = 0$ in this regime. The 68% of measurements converges to the standard deviation only for large values of $N$. The spread is $.32$ for the 76% and $.28$ for the 68%, which is both comparable to the middle 50%.

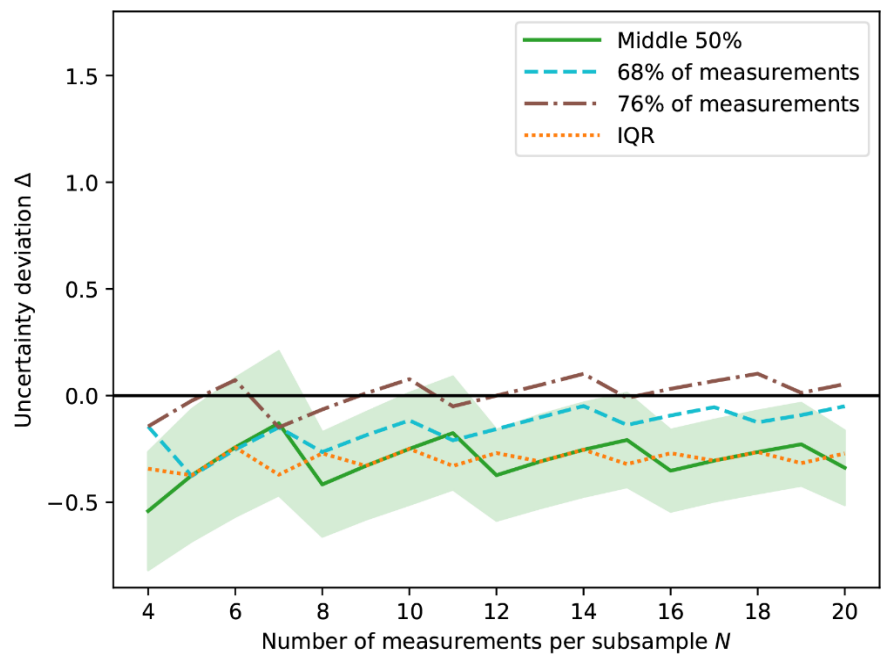

**Figure 4.** The mean values (lines) and standard deviation (shaded area, middle 50% only) of the uncertainty deviations of the different alternative uncertainty measures $u_i$ for different subsample sizes N. The horizontal black line indicates zero uncertainty deviation, corresponding to the standard deviation of the set. This figure shows the disregarded measures 68% of measurements, 76% of measurements, and the IQR compared with the middle 50%.

The IQR, unsurprisingly, oscillates around the middle 50% but has fewer kinks. This is due to the usage of the median in its calculation. The spread is $.22$, which is comparable to the middle 50%.

The lines of these three uncertainty measures, with the exception of the 76% of measurements line, fall well within the uncertainty region of the middle 50% line. This means that they give results of very similar quality. However, the complexity of these calculations is larger; the IQR requires medians to be calculated and the exclusion process for different percentages is more cumbersome. This renders these three uncertainty measures less useful than the middle 50%.

### 7.2 Implications for Teaching

The parallel increase in quality with complexity that **Table 1** shows, allows for a well-founded sequencing in uncertainty measures.

The usability of an uncertainty measure for teachers in experimental settings will depend on how well students can understand it. By taking an uncertainty measure with low complexity, students will get the best of both worlds: the determination is still done by hand, but simple mathematics will keep the focus on the understanding of the measurement result (Zangl & Hoermaier, 2017). The trade-off is that the quality with which the variability of the data is described is lower. Fortunately, many of the principles of statistical thinking and working with measurement uncertainties can be developed independently on the quality of a certain measure, that is, the concept of quantifying variance or comparing the variance of data sets (Torok & Watson, 2000, p.166).

To introduce the topic of measurement uncertainties, we recommend that teachers start with the min-max measure. This has the lowest complexity and thus provides an accessible way to introduce the topic of measurement uncertainties. Students have more room to evaluate their results, compare them, and think about the statistical nature of measurements. There are some practical materials available that use this definition of uncertainty (Hellwig et al., 2017; Kok & Boczianowski, 2021).

Still, this measure is highly susceptible to outliers, as shown in **Figure 3**, which should be discussed with students. This discussion of outliers will require students to think about the statistical nature of measurements and improve their overall understanding of what the measurement uncertainty describes.

Refining the measure to the exclude extremes measure is the next logical step. This measure has the same level of complexity, but offers a higher quality, especially in the case of eight repeated measurements.

The discussion of the relevance of repeated measurements can continue with students. Recording more measurements will inevitably lead to more outliers, rendering the exclude extremes measure also susceptible to outliers for a large number of repeated measurements. At this point, a converging measure like middle 50% can be introduced. Instead of 50%, one could also choose 68% or 76%– these percentages result in a slightly smaller deviation bias –but we feel that 50% will prove easier to determine for students during the step in which measurements are excluded from the calculation.

Ultimately, one will want to move towards using the standard deviation to describe the variability of the measurements. As an intermediate step, the MAD could be introduced. The MAD measure has some very auspicious characteristics. In contrast to the middle 50%, the MAD is smooth, converges quickly, and has the smallest spread of all uncertainty measures (see **Table 1**).

Introducing the MAD before standard deviation, as also suggested by Kader (1999), seems a logical step. The calculation of the MAD closely resembles the calculation of the mean value as well as the standard deviation but without the difficulty of squaring and taking the square root. The result –the mean uncertainty deviation– is an intuitive measure that is easier for high school students to interpret than the standard deviation. In scientific practice, the MAD is not very common. This is due to its horrible behavior when differentiating (Barlow, 1993, p.12). This is, however, a flaw that has no real consequences for high school practice.

As a last step in the series, the standard deviation can be introduced. This is the most complex measure and requires a more statistical approach for its interpretation due to the probabilistic nature of the Gaussian distribution which it is derived from. Therefore, it is advised to introduce this measure only in the upper school. It can easily tie in with the topic of the normal distribution, usually taught in the upper school. Additionally, it will prepare students for the formal scientific approach to calculate measurement uncertainties.

For a solid understanding of measurement uncertainties and data analysis, the topic of measurement uncertainties needs to be repeated in classrooms whenever appropriate (Munier et al., 2013). Only this way, real statistical thinking can be developed, which will also help with their appreciation of statistics later on (Casleton et al., 2014). We therefore suggest introducing the concept as early as possible and starting with a low complexity uncertainty measure fitting the mathematical capabilities of the students.

## Supplemental Material

The scripts to calculate and plot the results of this work can be found on GitHub: https://github.com/karel-kok/uncertainty_measures